\documentclass[namedreferences]{solarphysics}
\usepackage[pdfborder={0 0 0 },urlcolor=blue,breaklinks]{hyperref}
\usepackage[optionalrh,solaromanenum]{spr-sola-addons} 
\usepackage{graphicx}                    
\usepackage{color}                       

\ifx \arxivurl  \undefined \def \arxivurl#1{\href{http://arxiv.org/abs/#1}{\textsf{#1}}}\fi 
\ifx \doiurl    \undefined \def \doiurl#1{\href{http://dx.doi.org/#1}{\textsf{#1}}}\fi 
\ifx \adsurl    \undefined \def \adsurl#1{\href{http://adsabs.harvard.edu/abs/#1}{\textsf{#1}}}\fi 

\newcommand{\adv}{    {\it Adv. Space Res.}}

\newcommand{\aap}{    {\it Astron. Astrophys.}}

\newcommand{\apj}{    {\it Astrophys. J.}}

\newcommand{\mnras}{  {\it Mon. Not. Roy. Astron. Soc.}}

\newcommand{\solphys}{{\it Solar Phys.}}

\newcommand{\ssr}{    {\it Space Sci. Rev.}}

\begin{document}

\begin{article}

\begin{opening}

\title{On-line Tools for Solar Data Compiled at the Debrecen Observatory and their Extensions with the Greenwich Sunspot Data}

\author[addressref=aff1,corref,email={baranyi@tigris.unideb.hu}]{\inits{T.}\fnm{T.}~\lnm{Baranyi}}
\author[addressref=aff1,email={gylajos@tigris.unideb.hu}]{\inits{L.}\fnm{L.}~\lnm{Gy\H{o}ri}}
\author[addressref=aff1,email={ludmany@tigris.unideb.hu}]{\inits{A.}\fnm{A.}~\lnm{Ludm\'any}}

\address{Heliophysical Observatory, Konkoly Observatory, Research Centre for Astronomy and Earth Sciences, Hungarian Academy of Sciences, 4010 Debrecen, P.O. Box 30, Hungary}

\runningauthor{T. Baranyi {\it et al.}}
\runningtitle{On-line tools for Debrecen and Greenwich data}

\begin{abstract}
The primary task of the Debrecen Heliophysical Observatory (DHO) has been the most detailed, reliable, and precise documentation of the solar photospheric activity since 1958. This long-term effort resulted in various solar catalogs based on ground-based and space-borne observations. A series of sunspot databases and on-line tools were compiled at DHO: the Debrecen Photoheliographic Data (DPD, 1974--), the  dataset based on the {\it Michelson Doppler Imager} (MDI) of the {\it Solar and Heliospheric Observatory} (SOHO) called SOHO/MDI--Debrecen Data (SDD,  1996--2010), and the dataset based on the {\it Helioseismic and Magnetic Imager} (HMI) of the {\it Solar Dynamics Observatory} (SDO) called SDO/HMI--Debrecen Data (HMIDD, 2010--).  User-friendly web-presentations and on-line tools were developed to visualize and search data. As a last step of compilation, the revised version of Greenwich Photoheliographic Results (GPR, 1874--1976)  catalog was converted to DPD format, and a homogeneous sunspot 
database covering more than 140 years was created. The database of images for the GPR era was completed with the full-disc drawings of the Hungarian historical observatories \'Ogyalla and Kalocsa (1872--1919) and with the polarity drawings of Mount Wilson Observatory. 
We describe the main characteristics of the available data and on-line tools.

\end{abstract}

\end{opening}

\section{Introduction}

The most important observable phenomena for the investigation of the solar activity are the surface magnetic fields, in particular the longest observed features: the solar active regions, sunspots, and sunspot groups. For this reason reliable documentation of the sunspot activity is of basic importance for the understanding of the solar dynamo and also for any other components of the solar activity. Two parallel approaches have been established in solar research. One of them provides a single value for each day with no regard for structural properties of sunspot groups. This approach resulted in the series of International Sunspot Number (ISSN: \citealt{Clette07}) and Group Sunspot Number  \citep{Hoyt}. They are based on full-disc solar drawings, which can be extended back to the first telescopic observations in 1610. These data are indispensable for the long-term studies of solar activity of the last four centuries. The ISSN is widely used as the main index of solar activity; this is the reason for the 
recent cooperative efforts to ensure the homogeneity of this dataset \citep{Clette14}. 

The task of the other approach is the recording of the observable properties of all active regions, primarily their sizes and positions. These databases are based on white-light full-disc photographic observations, and they play an important role in the study of spatial and temporal distributions of sunspots and active-region development. The first detailed sunspot catalog was the Greenwich Photoheliographic Results (GPR: \citet{RGO}) described by \citet{Willis13a, Willis13b, Willis16} and \citet{Erwin13} recently. The volumes of GPR contain the position and area data of all observable sunspot groups on a daily basis. In the first decades, the data of some (not all) individual sunspots were also published, and it also contained white-light facular data until 1955. 

When the GPR program was terminated, the International Astronomical Union charged the Debrecen Heliophysical Observatory (Dezs\H o, 1982) with the continuation of the program from 1977 onwards. The present article describes the current state and services of the Debrecen sunspot-data programs. The core program is the Debrecen Photoheliographic Data (DPD), the first sunspot catalog containing the position and area data of all observable sunspots and sunspot groups on a daily basis; this is the formal continuation of the GPR with higher complexity.  The DPD has recently reached the total coverage of the post-GPR era, and it has been unified with the GPR. Now the GPR and the DPD constitute a homogeneous dataset with three overlapping years and all of the on-line tools of DPD have been extended for the GPR.  The DPD team also extended the catalog work to space-borne full-disc continuum observations. This team created sunspot and facular datasets based on the {\it Michelson Doppler Imager} (MDI) of the {\it Solar 
and Heliospheric Observatory} (SOHO) called SOHO/MDI--Debrecen Data (SDD,  1996--2010), and the datasets based on the {\it Helioseismic and Magnetic Imager} (HMI) of the {\it Solar Dynamics Observatory} (SDO) called SDO/HMI--Debrecen Data (HMIDD, 2010--2014). The space-borne catalogs are even more detailed databases containing the magnetic-field data of all spots with a temporal resolution of ${\approx}$ 1-1.5 hours. The included magnetic data greatly extend the possibilities for investigations because of the higher cadence and the distinction of leading and following polarities.

\section{Debrecen Photoheliographic Data}

At present, the most detailed ground-based  catalog is the DPD, providing area and position data for each observable sunspot on a daily basis along with images of sunspot groups, full-disc scans and magnetograms. The DPD is mainly compiled by using white-light full-disc observations taken at DHO and its Gyula Observing Station with an archive containing more than 150,000 photoheliograms observed since 1958. Observations of a number of other observatories around all of the world help in making the catalog complete.

The numerical part of the DPD contains area and position of each spot, the total areas and the mean positions of the sunspot groups, and the daily sums of the area of groups. These three kinds of data relating to spots, sunspot groups, and daily sums are organized into three kinds of rows of data.  The first character of the row indicates the type of the row. If the first letter is $s$, it means that the row contains data for a spot. If the first character is $g$, the row contains the total areas and the mean positions of a sunspot group. If the first character is $d$, the row contains daily data. The following data are available for each spot: time of observation, the NOAA number of its group, the measured (projected) and the corrected (for foreshortening) areas of umbrae [$U$] and the whole spot [$WS$, formerly $U+P$], latitude [$B$], longitude [$L$], distance in longitude from the central meridian [$LCM$], position angle [$P$], and distance from disc center [$r$] expressed in solar radii. Several kinds of 
numerical data are presented in ASCII files: yearly tables for daily sums of area data; time series of daily data (rows $d$ only); tables containing the whole area of sunspot groups and their mean position data (rows $g$ only);  tables for the sunspot area and position data (rows $s$ only); combined datasets containing all three kinds of rows.

The DPD contains of all the observable sunspots, and it also contains data for umbrae. If there is a darker part within a spot, this part can be identified as an umbra based on its intensity exclusively. The corrected area of the whole spot [$WS$] and the umbral area [$U$] are measured in millionth of solar hemisphere [msh]. If the observed area is smaller than 0.5 msh, it is indicated with 0 msh at the measured position. A zero {$U$} may mean that the derived $U$ is smaller than 0.5 msh, or the observable structure of the spot does not allow one to identify any internal pattern. If an umbra is identified within a spot, the position of the umbra is published in the row $s$. Otherwise, the centroid of the spot determines the position of the spot. If there is more than one umbra within a penumbra, this fact is also published. In such a case, the $WS$ is published in the row of one of these umbrae and the rows of the other umbrae contain the ordinal number of this umbra with negative sign, which indicates that 
the given umbra shares a penumbra with the umbra of the ordinal number indicated with negative sign.
({\it E.g.} as can be seen in Figure 3, the umbra Number 2 shares the penumbra with umbra Number 1. This fact is indicated in such a way that there is -1 in the columns of $WS$ in the row of the umbra Number 2. It is indicated in a similar way that the umbra Number 4 shares its penumbra with lots of other umbrae.)

Scans of sunspot groups are appended showing the spots numbered as in the numerical dataset.  
Full-disc white-light images and magnetic observations are appended to provide the morphological and polarity information available concerning the sunspots. All of the data and images are accessible by ftp to provide an easy bulk download, but the entire material is also provided in a user-friendly interactive graphical presentation. The daily page of the on-line presentation of the data contains a schematic full-disc drawing created from the spot data of DPD and a magnetic observation. Below the schematic full-disc drawing, there is a link to open the jpg version of the original full-disc white-light observation in a pop-up window. All of the information on a sunspot group can be reached by clicking on the group number. The days can be surveyed by turning the pages both at the daily pages and at the group pages. Figures 1 and 2 show the full-disc images available for a day at the website of DPD and Figure 3 shows an example for a page of sunspot data of DPD here.

There is also an on-line MySQL query at the website of the catalogs, which makes possible a quick and easy selection of the numerical data (Figure 4).

\begin{figure}
\begin{center}
  \includegraphics [width=12cm]{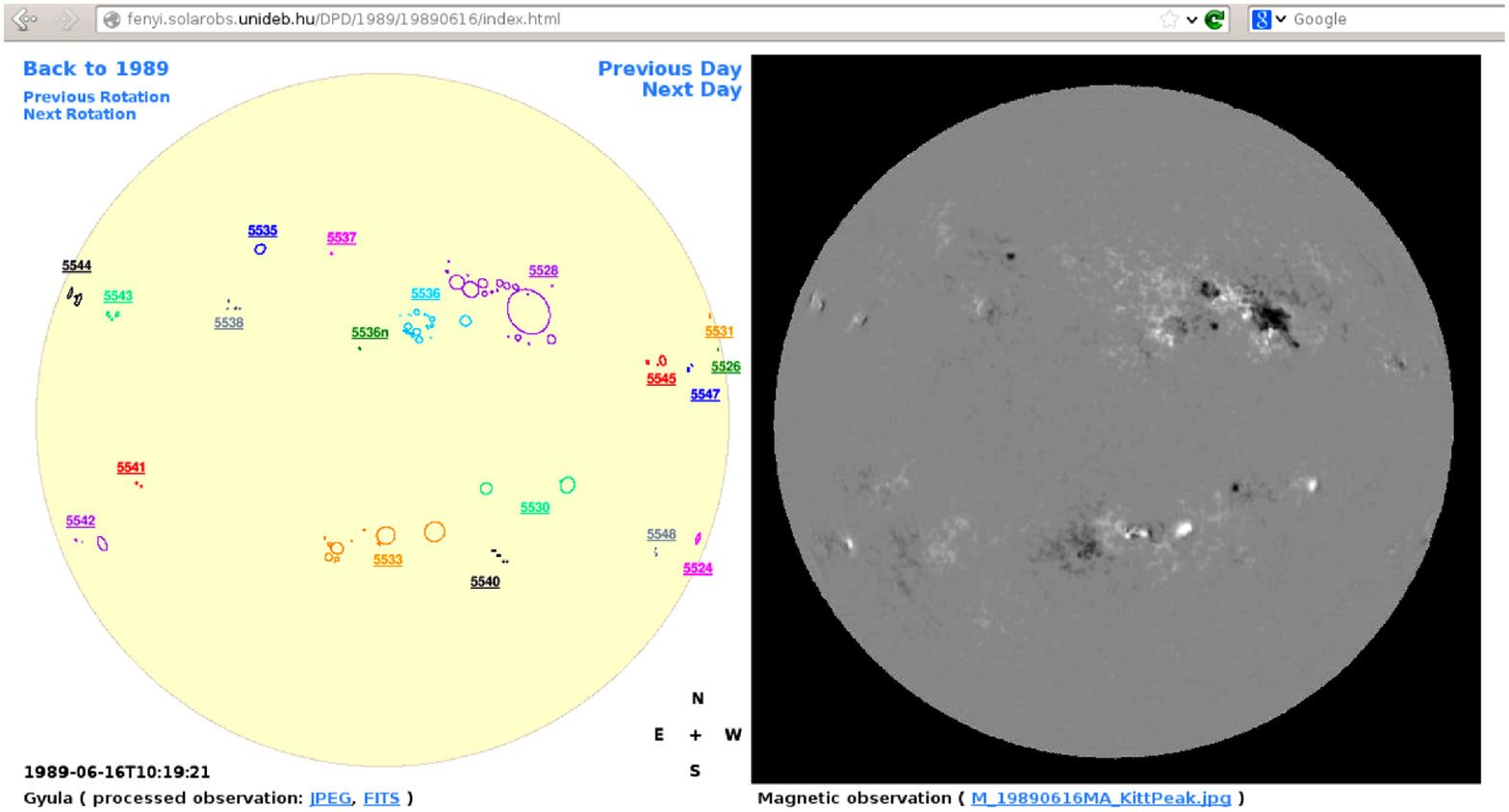}
\caption{Example of a web page showing the available information for a given date.  Left panel: Schematic drawing of sunspots and sunspot groups visualizing the data content of DPD. It is created from the position and area data of spots, derived from the observation indicated at the bottom of the panel. The spots (and pores) are represented by ellipses visualizing approximately the spot roughly as a projection of a circle on a sphere onto a plane. The centroid of the ellipse is at the position of the visible centroid of the spot. The area of the ellipse is the projected whole spot area [$WS$]. Right panel: Magnetogram or polarity drawing available for the given day.}
\end{center}
\end{figure}

\begin{figure}
\begin{center}
  \includegraphics [width=12cm]{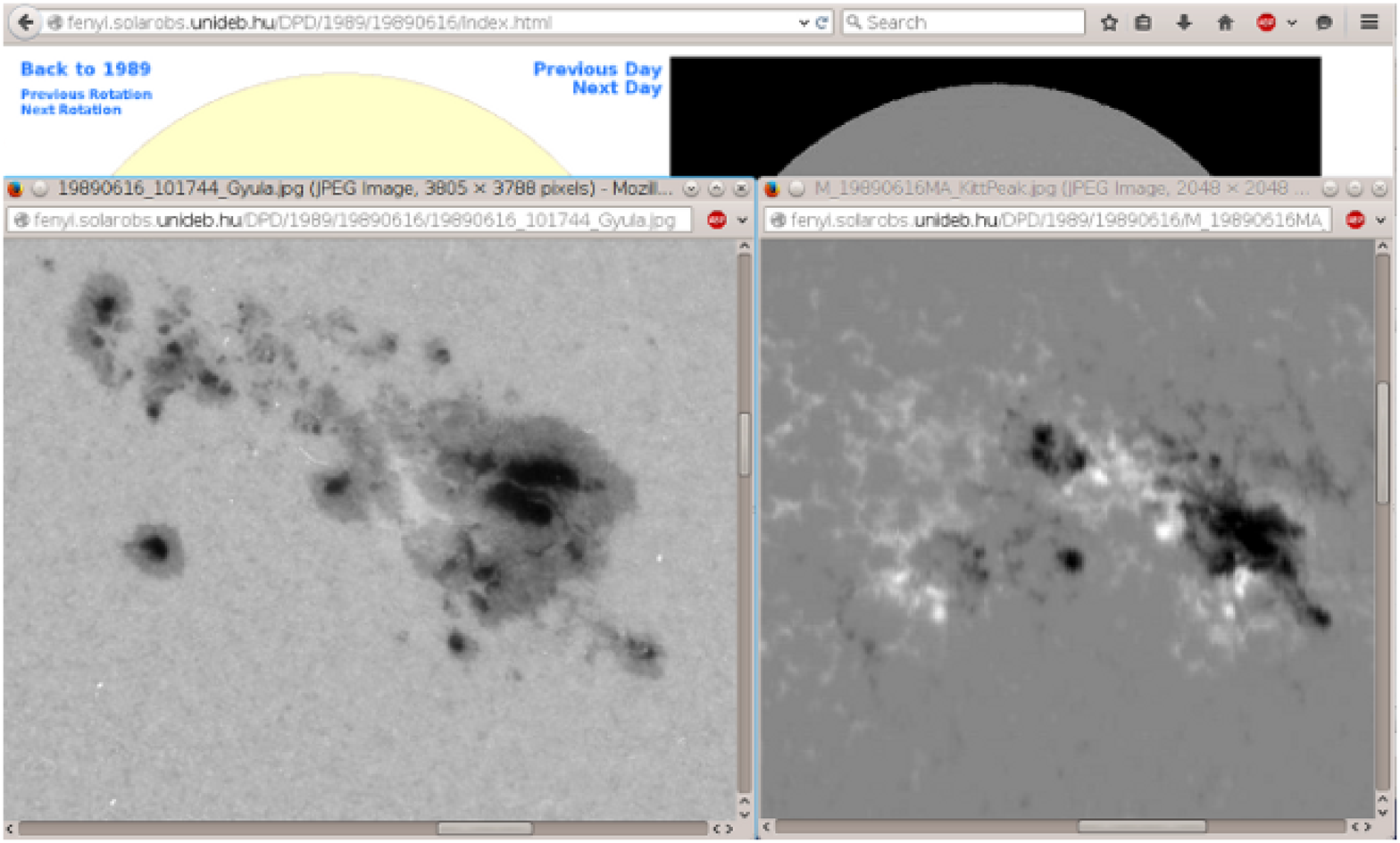}
\caption{ Example for pop-up windows opening after clicking on the links below the images shown in Figure 1. The full-disc white-light image and magnetogram can be seen by scrolling the side bars of the pop-up windows. Here an example for a possible set up of side bars is shown where the pop-up windows are centered on the AR NOAA 5528. The spatial scale of the image in the pop-up window is the same as the spatial scale of the original observation or its scanned version. The images gathered from the various archives may have different spatial scales.}
\end{center}
\end{figure}

\begin{figure}
\begin{center}
  \includegraphics [width=7cm]{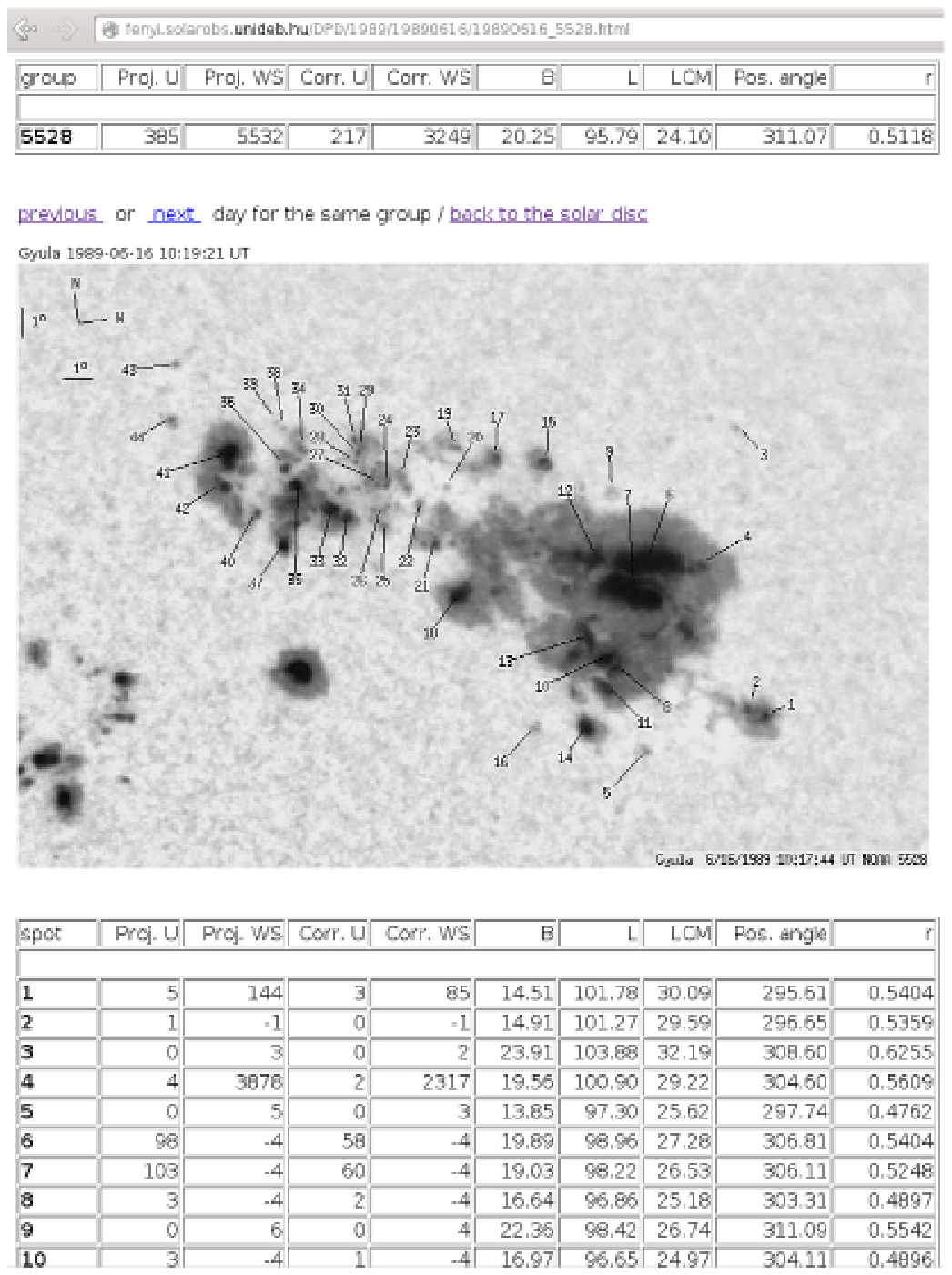}
\caption{ Example for a page opening after selecting the link at AR NOAA 5528 in Figure 1.  The spatial scale of the image of sunspot group is shown by line segments of one heliographic degree long in the upper left corner. All of the numerical data of the numbered features of the sunspot group can be seen in the table below the image.}
\end{center}
\end{figure}

\begin{figure}
\begin{center}
  \includegraphics [width=10cm]{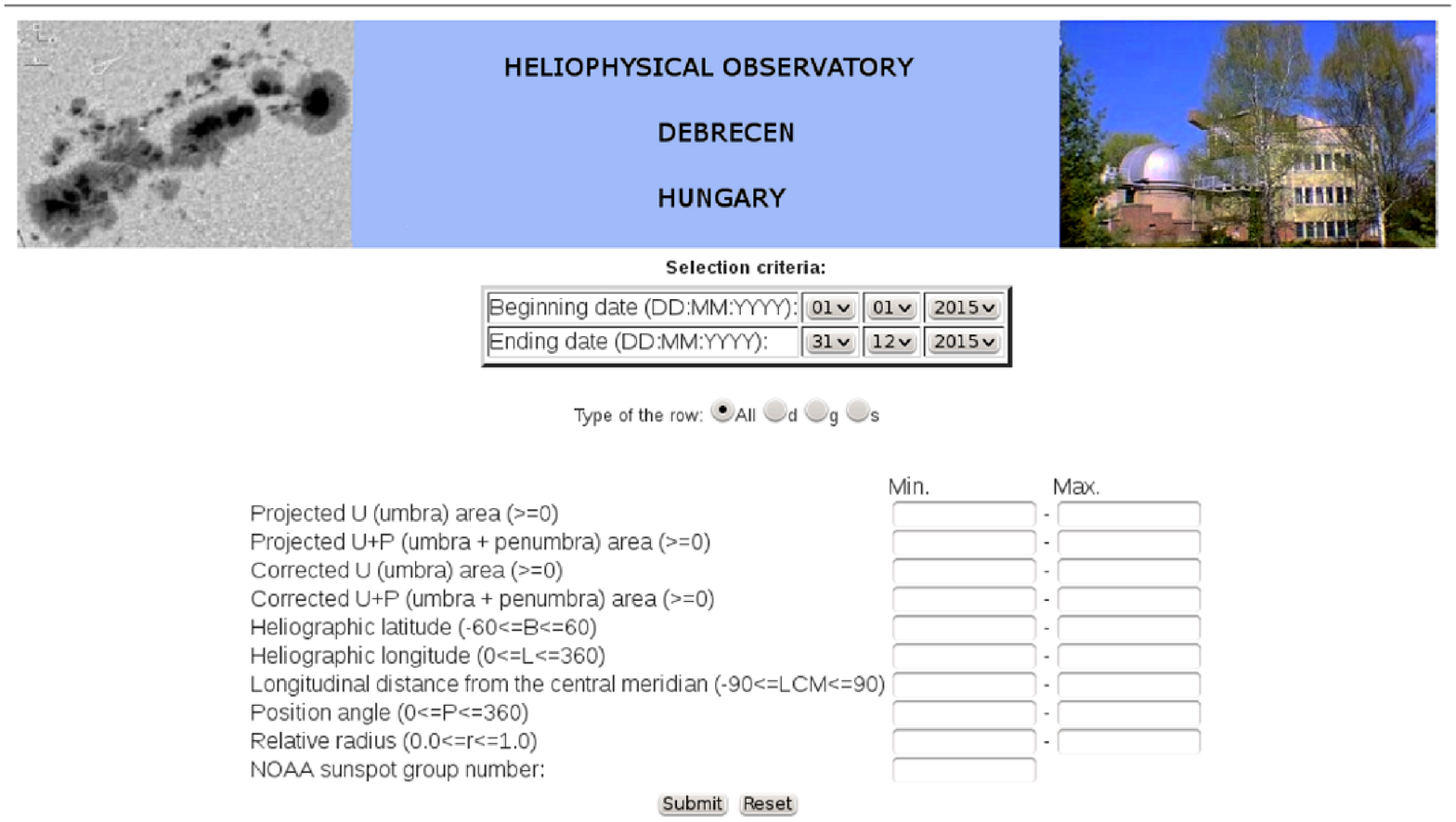}
\caption{Page of the MySQL query of DPD to select the data by applying selection criteria at {\sf fenyi.solarobs.unideb.hu/test/query/} }
\end{center}
\end{figure}

The DPD is available starting in 1974. Some of its volumes are still in a preliminary format and need a further quality check. The instrumental background for DPD  catalog compilation has changed over the years \citep{Gea05, Gyori11, Gea16}. The software package called Sunspot Automatic Measurement (SAM) (Gy\H ori 1998, 2005) was developed to handle the scans of the ground-based 
photographic observations. After that, it was further developed to handle the ground-based and space-borne CCD FITS images. 
The rate of space-borne observations included in the DPD increased in recent years. However, the ground-based observations remain essential to contribute to the completeness of the whole material. 
Detailed analysis of the precision of the data can be found in the articles by \citet{Baranyi01, Baranyi13, Gyori12, Gea16}.

\section{SOHO/MDI--Debrecen Data (SDD)}

The SOHO/MDI-Debrecen Data (SDD)  catalog is based on the SOHO/MDI continuum intensity images and magnetograms \citep{Scherrer95}. This  catalog is similar to that of DPD in its data format, image products, and on-line tool but the temporal cadence is about one hour depending on the availability of MDI observations. The novelty is that the SDD contains magnetic information because the SAM is suitably modified to determine the mean line-of-sight magnetic field of umbral and penumbral parts of spots from the quasi-simultaneous magnetogram. 

The software automatically finds the sunspots in the solar-disc images of 1024{$\times$}1024 pixels, it draws their internal (umbra) and external (penumbra) contours, and it determines their positions, areas, and mean magnetic-field strength. The full-disc version of the SDD  catalog (fdSDD) is based on these data. The spots are numbered by the computer program on the basis of their longitude, and they are not assigned to sunspot groups. In the next step, the arrangement of spots into sunspot groups was made automatically by using the sunspot-group data of the pre-existing DPD  catalog,  finally, the arrangement was checked and corrected by human assistance.  The procedure of checking and improving is time consuming so it is not yet complete. In spite of its partially preliminary state, this  synoptic dataset makes it possible to investigate the internal dynamics and evolution of the active regions with high time resolution. 

The unique level of detail of SDD can be demonstrated with a series of images for the sunspot group NOAA 10486 in Figure 5. This group produced a powerful X17.2-class flare on 28 October 2003. Figure 5 shows the last three sets of data in SDD before this flare. The schematic polarity drawing of the sunspot group in the middle column of Figure 5 illustrates the data content of SDD. Each image in the middle is created from the position and area data of spots derived from the observations alongside it. The spots (and pores) are represented by ellipses outlining approximately the spot roughly as a projection of a circle on a sphere onto a plane. The centroid of the ellipse is at the position of the visible centroid of the spot. The area of the ellipse is the projected whole spot area. The projected umbral area is represented by a smaller ellipse within the ellipse of the spot. The colors of the ellipses of umbra and penumbra show the polarities of their mean magnetic field derived from the magnetogram (light-
gray penumbra and white umbra correspond to positive polarity, while dark-gray penumbra and black umbra correspond to negative polarity). At least a small part of the spot is always colored in white or black (even if $U$=0) to emphasize the polarity information. In the case of several umbrae in common penumbra, there is an umbral ellipse for each umbra at its position. In the case of mixed polarities, the colors of umbra and penumbra show opposite polarities. The amount of data for this group is more than two orders of magnitude larger than that of DPD; the SDD contains more than 10,000 independent  (non-redundant) position, area, and magnetic data of spots in this group for this day (15 images per day; about 120 spots in this group per image; two position data, two area data, and two magnetic data per spot).

\begin{figure}
\begin{center}
  \includegraphics [width=12cm]{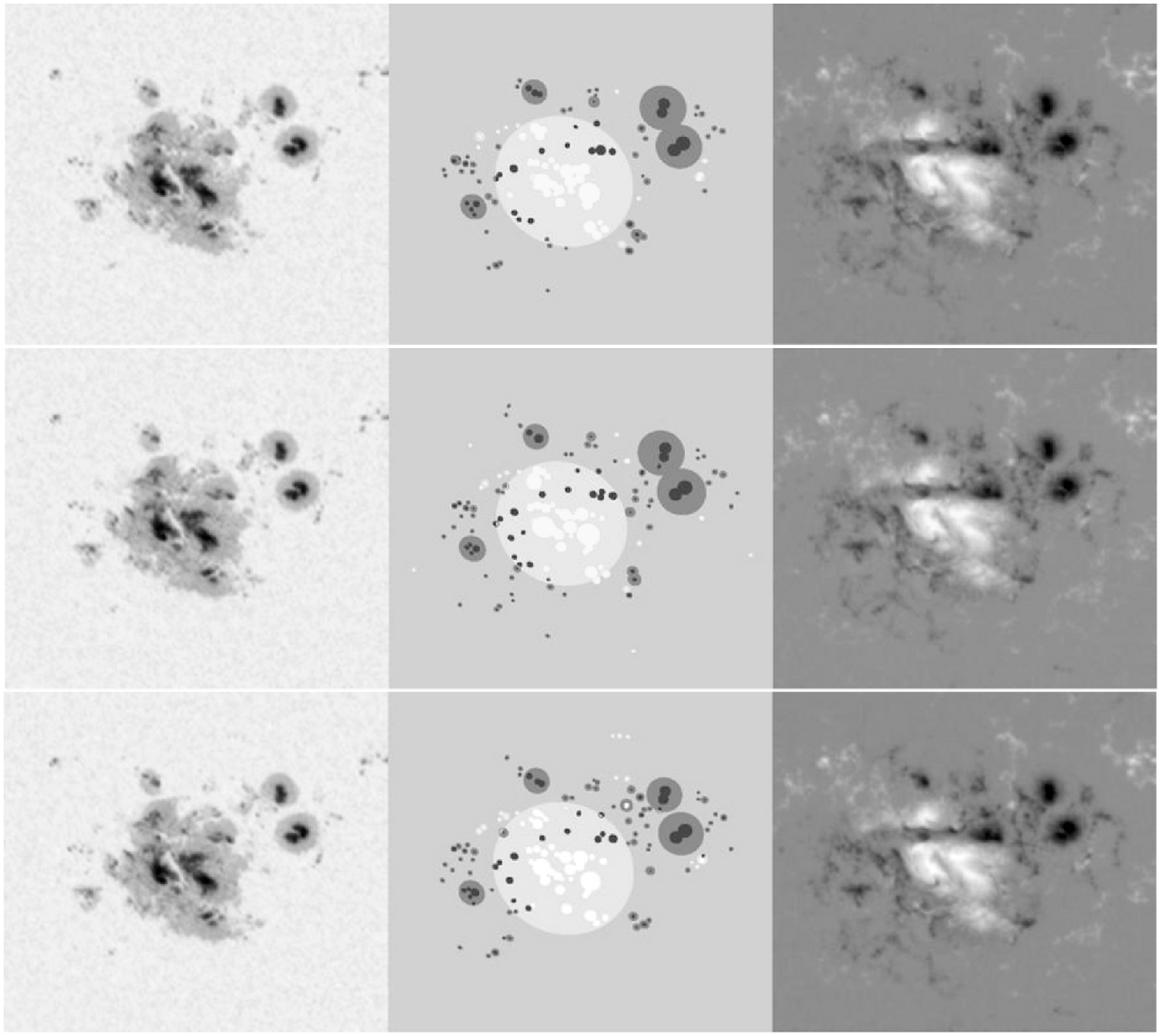}
\caption{The unique level of detail of the SDD sunspot  catalog can be seen if one compares the schematic drawing created by using only the published data (middle column) with the original continuum image (left column) and magnetogram (right column) of MDI.  The figure shows the images for NOAA 10486 on 28 October 2003.  The time of observation in the first row 06:23:33 UT, in the second row 07:59:33 UT, in the third row  09:35:33 UT. The spatial extent of the images is about 22{$\times$}19 heliographic degrees.}
\end{center}
\end{figure}

\section{SDO/HMI--Debrecen Data (HMIDD)}

For the HMIDD, the data structure and the on-line tools are very similar to those of SDD. The difference comes from the fact that the spatial resolution of HMI is larger than MDI. The 4096{$\times$}4096 pixels high-quality images allow measuring much smaller features with higher precision in the images. On one hand, this results in large data files with somewhat different format from that of DPD and SDD. On the other hand, it results in  sunspot-group images where the numbers and indicating lines of spots are too crowded.

Because of the large amount of data, the presentation of the data with a tool showing detailed information has a greater importance. The website of HMIDD is extended with extra pages at the links ``See sunspots with tool" below the images in the pages of sunspot groups. This simple tool is useful to study positions and polarities of sunspots by completing the sunspot-group images in that page with additional images in high resolution and polarity drawings.
In the upper panel of these pages, the white-light image or the magnetic observation of a sunspot group can be seen depending on the viewer choice by the radio button on the right while the lower panel shows the schematic polarity drawing of sunspot group reconstructed from the HMIDD data.
If one moves the mouse cursor over a sunspot (or umbra) near its centroid in the upper or lower panel, the actual related data row of that sunspot will pop up on the row between the upper and lower panels as the data of spot (umbra) No. 26 can be seen in Figure 6.
To help browsing data, the full-disc schematic drawings reconstructed from the data are also published ({\it e.g.}  Figure 7 is created from position, area, and magnetic-field strength data of 566 sunspots derived from the continuum image and magnetogram for 2014--07--06T20:00:41UT). 

The huge amount of data in the HMIDD requires that the assignment of spots to sunspot groups is processed with an automatic method exclusively. This method is mainly based on the information on sunspots and sunspot groups listed in DPD. That spot, which is not included into DPD, is assigned to the closest group of DPD if the distance is smaller than five heliographic degrees. If there are no DPD groups within this distance, the spot or the cluster of nearby spots is assigned to a newly created sunspot group with a name created from an existing NOAA active-region number by adding a previously unused letter to it. 

The quick-look version of DPD is also based on the HMI observations. To provide daily sunspot data as soon as possible, one HMI image per day is regularly evaluated starting the workdays with this task and publishing the data within a few hours.   

\begin{figure}
\begin{center}
  \includegraphics [width=12cm]{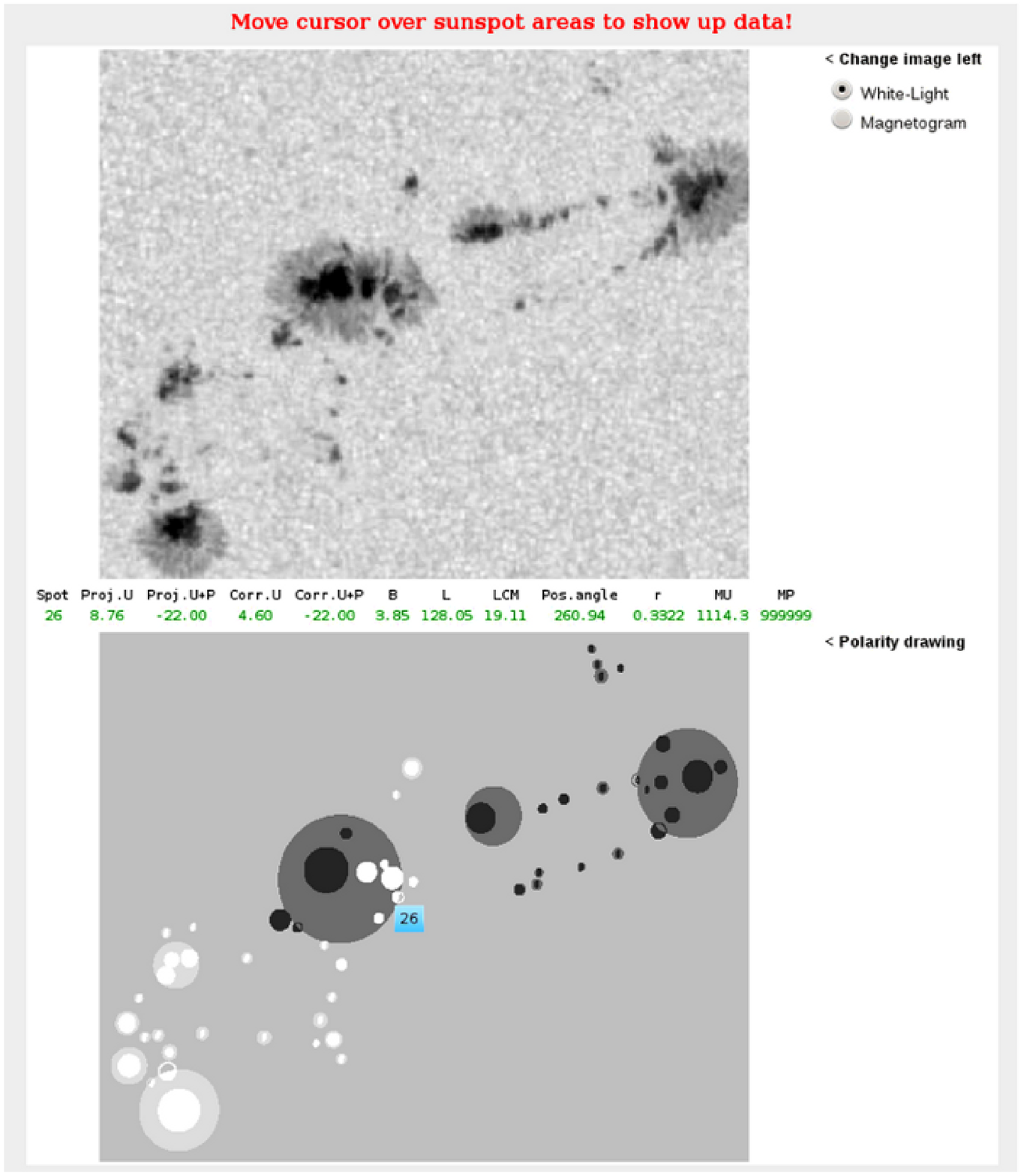}
\caption{Sample of page available at the link ``See sunspots with tool" below the image in the page of sunspot group, {\it e.g.} at {\sf fenyi.solarobs.unideb.hu/ESA/2012/20120903-005906.80/20120903-005906.80\_11560.html}. 
Upper panel: White-light image of sunspot group. It can be seen that this image can be replaced with the magnetic observation if the user chooses  that option button on the right. Lower panel: Schematic polarity drawing created from the data as it is described in the case of Figure 5. If one moves the mouse cursor over a sunspot (or umbra) near its centroid in the upper or lower panel, the actual related data row of that sunspot will pop up on the row between the upper and lower panels as the data of spot(umbra) Number 26 can be seen in this figure. The spatial extent of these images is about 11{$\times$}9 heliographic degrees in this case. Before creating these images, the HMI observations are enlarged from 4096{$\times$}4096 pixels to 6600{$\times$}6600 pixels.}

\end{center}
\end{figure}
\begin{figure}
\begin{center}
  \includegraphics [width=10cm]{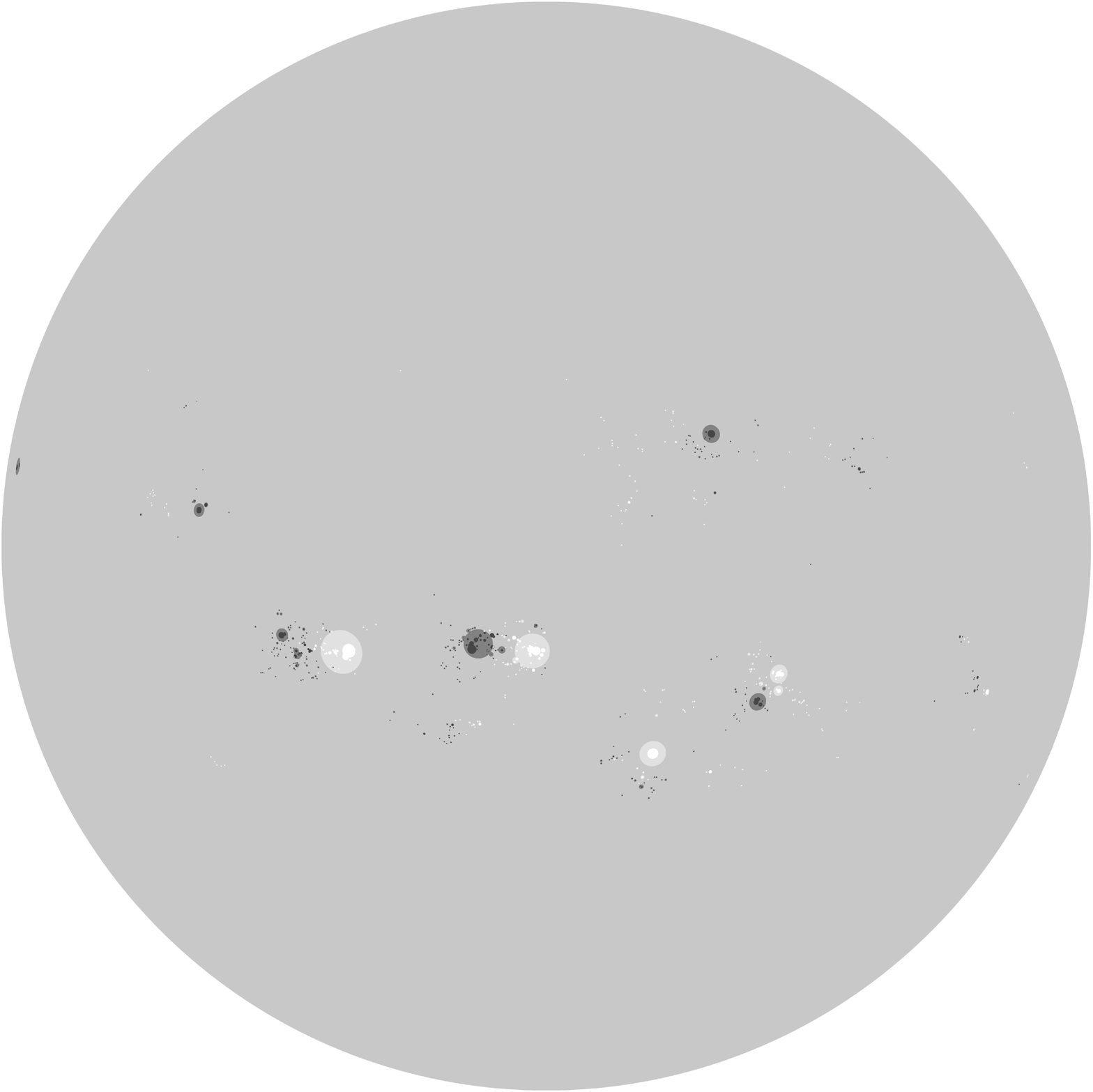}
\caption{Full-disc schematic drawing created from HMIDD data for 2014-07-06T20:00:41UT as it is described in the case of Figure 5. The images of this type are available at {\sf ftp://fenyi.solarobs.unideb.hu:2121/pub/SDO/images/Polarity\_drawings/}.}
\end{center}
\end{figure}

\section{Catalogs of Sunspot Group Tilt Angles}

The direction of the line connecting the leading and following portions of a bipolar sunspot groups is usually tilted with respect to the solar Equator \citep{Hale}. The white-light images only allow a simple method for the estimation of this angle but more reliable tilt-angle data can be determined by distinguishing between the polarities of spots.

The tilt angle is traditionally defined to range between $\pm 90^{\circ}$ and to be positive if the absolute value of the heliographic latitude of the leading part is smaller than that of the following part but other definitions are also possible (see, {\it e.g.}, \citealp{Li} and \citealp{McClintock14}). Considering the diagnostic importance of this angle, all catalogs of DHO have appendices containing the tilt angles of the sunspot groups. The DPD lacks magnetic information; in this case the method of \citet{Howard} had to be used, in which area-weighted positions of the umbrae of leading and following portions were derived in the portions located to the West and East of the area-weighted centroid of the entire sunspot group. This procedure has also been carried out by using the whole spot area as the weight. These resulted in two sets of columns of data. The novelty of SDD and HMIDD tilt-angle datasets is that they are based on two new different tilt definitions including the information of magnetic 
polarity of 
spots in addition to the traditional data. In this cases, the tilt angles also range between $\pm 90^{\circ}$ degrees but it is the polarity of the spot that determines its leading or following role instead of the position with respect to the centroid of the group.  In the case of space-borne datasets, we have four tilt data values for each group, which makes it possible to select the most unambiguous cases in which these values are closer to each other than a pre-selected criterion. These tilt-angle data and their methodology are described in detail by \citet{Baranyi15}. Figure 8 shows the on-line query available for DPD and SDD tilt data showing the options for selection criteria.

\begin{figure}
\begin{center}
  \includegraphics [width=10cm]{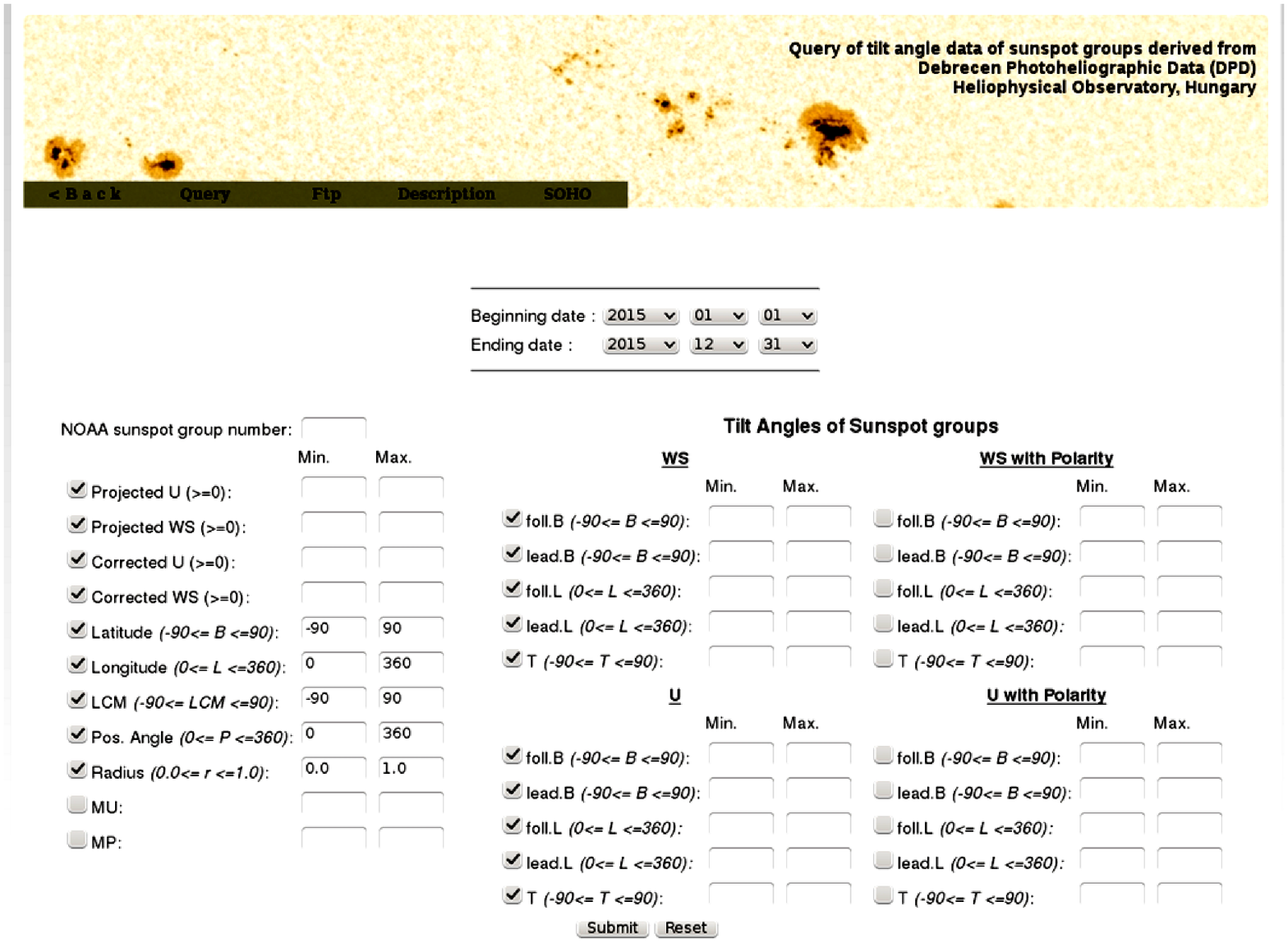}
\caption{Web page of the query for tilt-angle data showing the available selection criteria.
The queries for tilt angles are available at
{\sf fenyi.solarobs.unideb.hu/test/tiltangle/dpd/} and {\sf fenyi.solarobs.unideb.hu/test/tiltangle/sdd/}
}
\end{center}
\end{figure}

\section{Catalogs of White-Light Faculae}

The GPR contained facular data besides the spot data but the DPD did not include this additional task. However, the quality of the space-borne continuum images allowed the derivation of detailed facular data in a similar way to that in which spot data are measured. 
The full-disc facular data both in SDD and HMIDD have a similar format to that of full-disc sunspot data. The main difference is that the columns of umbral-area data are filled with zeros. The first column of magnetic data contains the line-of-site magnetic-field value at the brightest pixel within the facular contour in the intensity image.
Concerning the comparison of SDD and HMIDD facular data, see \citet{Gyori12}. Figure 9 shows an example for a page showing facular data.

\begin{figure}
\begin{center}
  \includegraphics [width=12cm]{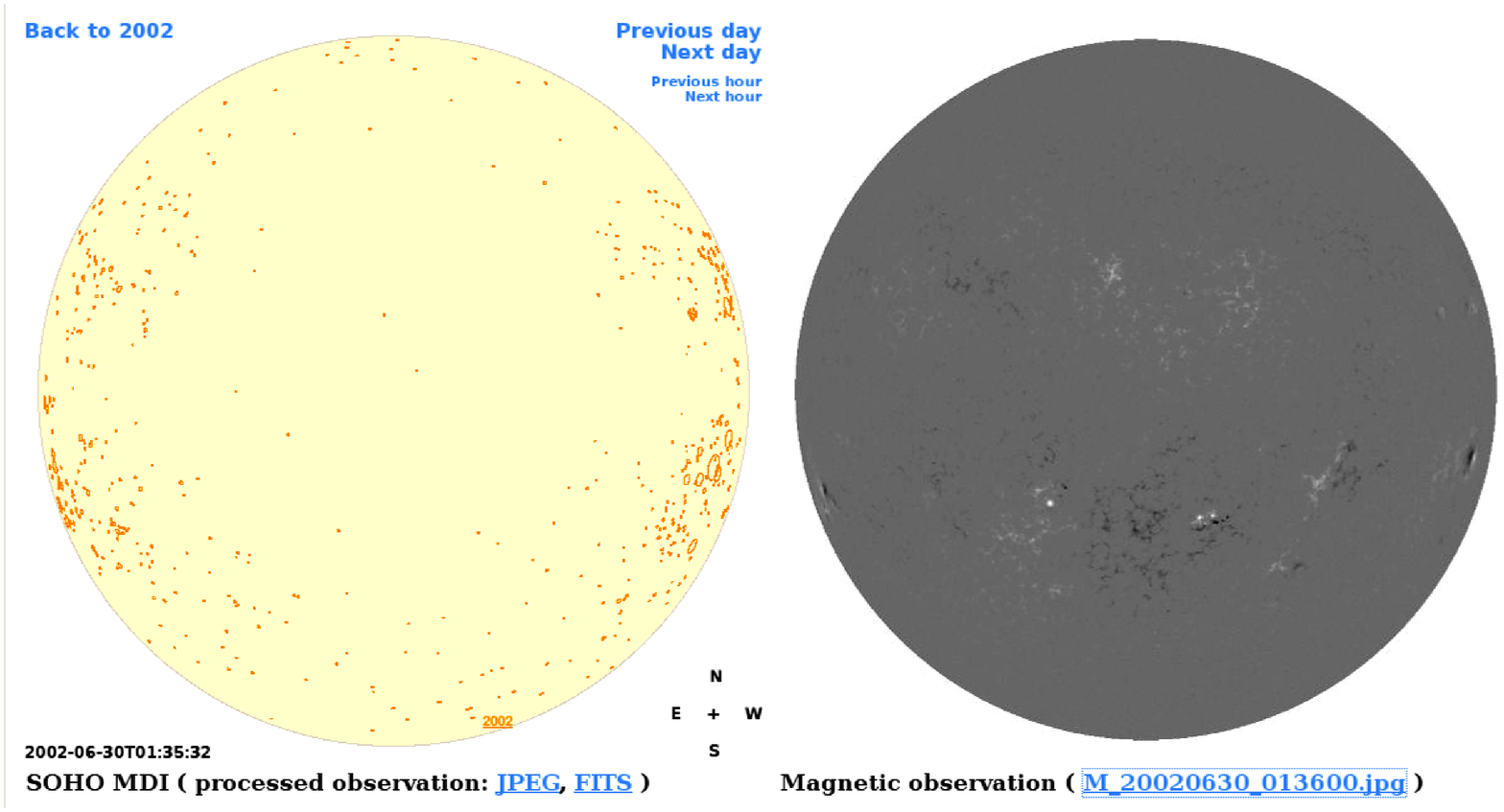}
\caption{Example for a web page visualizing facular data available at {\sf fenyi.solarobs.unideb.hu/SDD/faculae/2002/20020630-013532/}}
\end{center}
\end{figure}

\section{Hungarian Historical Solar Drawings (HHSD)}

 The DHO hosts solar-image databases containing heritages of two former Hungarian observatories. The first set of drawings was observed between 1872 and 1891 at the \'Ogyalla Observatory (now Hurbanovo, Slovakia) founded by Mikl\'os Konkoly-Thege (1842--1916) \citep{VK99}. The other set of solar drawings was observed at the Haynald Observatory in Kalocsa between 1880 and 1919 \citep{Toth02}. The  full set of drawings is available at the site of Hungarian Historical Solar Drawings ({\sf fenyi.solarobs.unideb.hu/HHSD.html}), but the selected images are also included in an interactive on-line presentation combined with the data of GPR. These drawings, which are rich in details, help us to reveal the structure of sunspots and sunspot groups in those years when there are no photographic images available. The drawings may also contain information for faculae drawn with colors different from those of spots. In a few cases, the features above the photosphere may also be seen at the limb observed by using a
spectrohelioscope. Two examples for these 
drawings can be seen in 
Figure 10.   
\begin{figure}
\begin{center}
  \includegraphics [width=12cm]{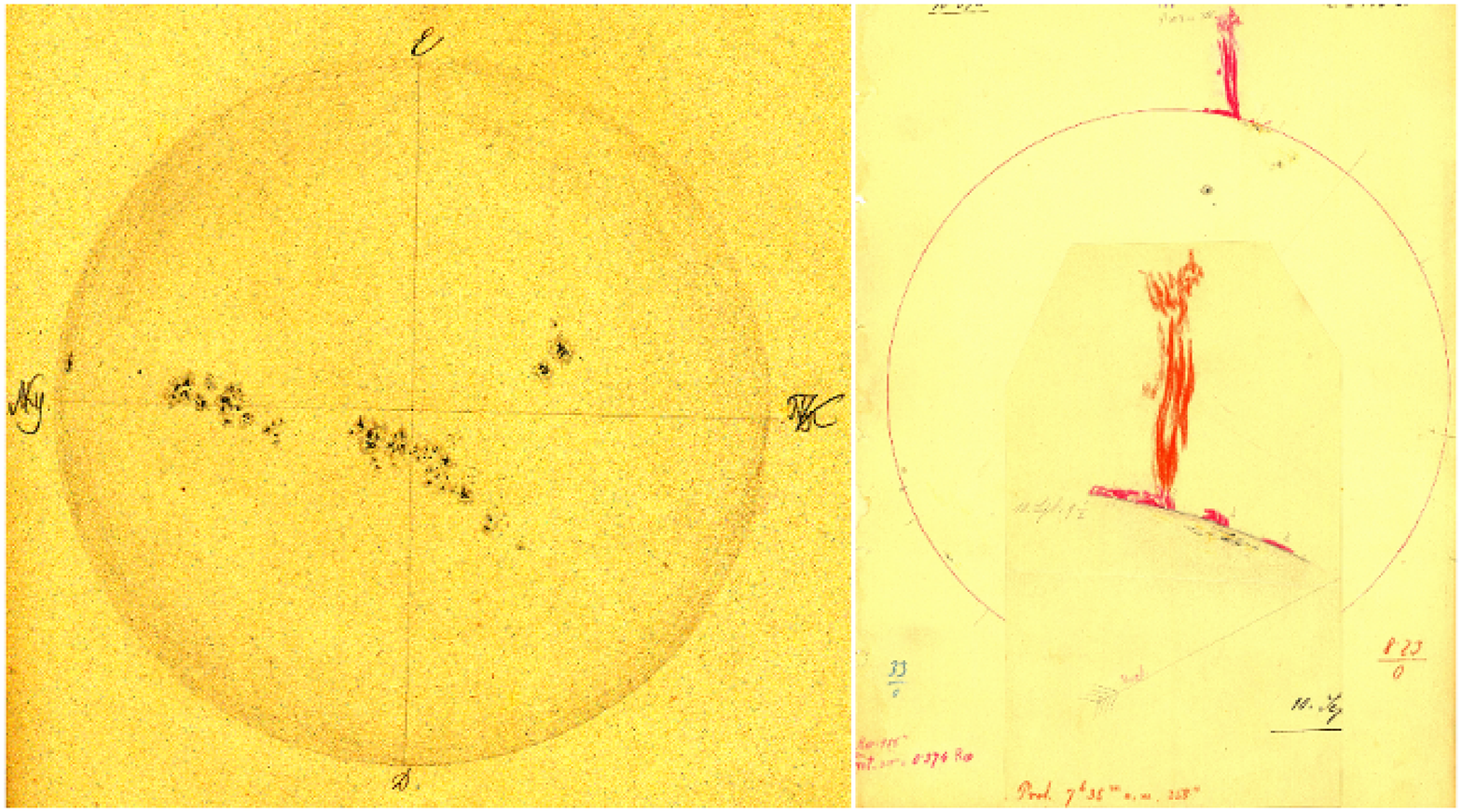}
\caption{Two examples for full-disc drawings in HHSD. Left panel: Observation taken at \'Ogyalla on 22 November, 1872 (East-West side-reversed). Right panel:  Observation taken at Kalocsa on 10 September, 1891 with an enlarged inset. The full set of HHSD is available at {sf fenyi.solarobs.unideb.hu/HHSD.html}. }
\end{center}
\end{figure}

\section{GPR Datasets and Their Revision}

The GPR contains the heliographic coordinates and total areas of the sunspot groups on a daily basis in each year between 1874 and 1976. In the beginning, they also included the data of some individual spots but later this was abandoned. Thus, only the group data were suitable to produce a homogeneous dataset for the whole GPR era.

The Greenwich books are available in pdf format at the site of the UK Solar System Data Centre (UKSSDC: {\sf www.ukssdc.ac.uk/}). They have two types of sections of sunspot group data between 1874 and 1955: the ``ledgers" of sunspot groups and the daily ``measures" of sunspots and faculae. 
The basic dataset (1874--1976) consists of group data in files with extension ``gpr".  It is the electronic version of GPR published at NOAA National Geophysical Data Center (NGDC: {\sf www.ngdc.noaa.gov}) based on the ledgers. That database was compiled by Ward, USAF AFCRL, in the 1960s and later updated by Hoyt in the 1990s.
The additional dataset of Solar White Light Faculae contains detailed daily lists of sunspots (1874--1915) or daily lists of sunspot groups (1916--1955) based on the tables of ``measures". The summaries of total daily data are available for 1956--1976 with extension ``sum". Concerning the digitized datasets of GPR, see \citealt{Willis13a, Willis13b, Erwin13}. 

The method of revision  was mainly based on the comparison of basic files with the extension ``gpr" and the additional files called ``saf" (spots and faculae). These datasets were compared to eliminate or decrease the discrepancies because of the various typographic errors of the printed or digitized versions. The comparison of various types of published or derived position and area data resulted in lists of errors, which were usually corrected after checking them against the books, or they were recalculated from the data reckoned as correct. If the discrepancy could not be resolved by checking the books, the most probable error was corrected to achieve a consistency within the given threshold. In some cases, the correction was made after checking the data against the HHSD or Mount Wilson (MW) drawings.

The main goal of our work was to improve the ``gpr" files because that dataset was suitable to be unified with DPD. However, the ``saf" files were also corrected in many cases to achieve agreement between the two types of files. We list below the types of errors or problems detected during the quality check of the data and the method used to solve the problem:
If a redundant row was found in the ``gpr" file, it was deleted ({\it e.g.} the row was a duplication, or it contained false or deviating data for a group). If there was a group in the ``saf" with no pair in the ``gpr", the data of the missing group were entered into ``gpr". If the observational date of a given group was different in the ``saf" and in the ``gpr", the date was checked against the book and the erroneous file was corrected.  If the name of the group in the ``saf" differed from its name in the ``gpr", a similar method of correction was used. The name was different in a number of cases because the symbols (*, **, \#, \#\#) added to the group number in the book and in the ``saf" files were transformed to additional numbers in the original version of ``gpr". These symbols were transformed to letters a, b, c, d in the revised version of ``gpr" for the years 1874--1915. The names created from the Carrington rotation remained unchanged in the years 1956--1976. In this way, each group had 
a unique name in the revised ``gpr" dataset. We have also searched for outliers in the position and areal data. If the difference between the position of the group derived from the ``saf" and that in the ``gpr" was larger than one heliographic degree, we checked the error against the books. The same was made if the difference between the corrected area of a group in the ``saf" and  in the ``gpr" was larger than 5 msh. We also checked whether the projected or corrected total daily area data computed from ``gpr" data were different from those contained in ``sum" files. We corrected two types of errors after checking the internal consistency of the ``gpr" data. If the difference between the published heliographic coordinates $B$ and $L$ and those values of $B$ and $L$ that are derived from the polar coordinates $P$ and $r$ is larger than a radius-dependent threshold, the polar coordinates were recalculated from $B$ and $L$.
If the difference between the published projected area ($WS$ or $U$) and the projected area derived from the corrected area is larger than 10 msd and 10\,\%, the projected area was recalculated from corrected area and position data. By using this method, we achieved that a number of errors and outliers were filtered from the revised version of the GPR, and its level of internal consistency was increased.

\begin{figure}
\begin{center}
  \includegraphics [width=12cm]{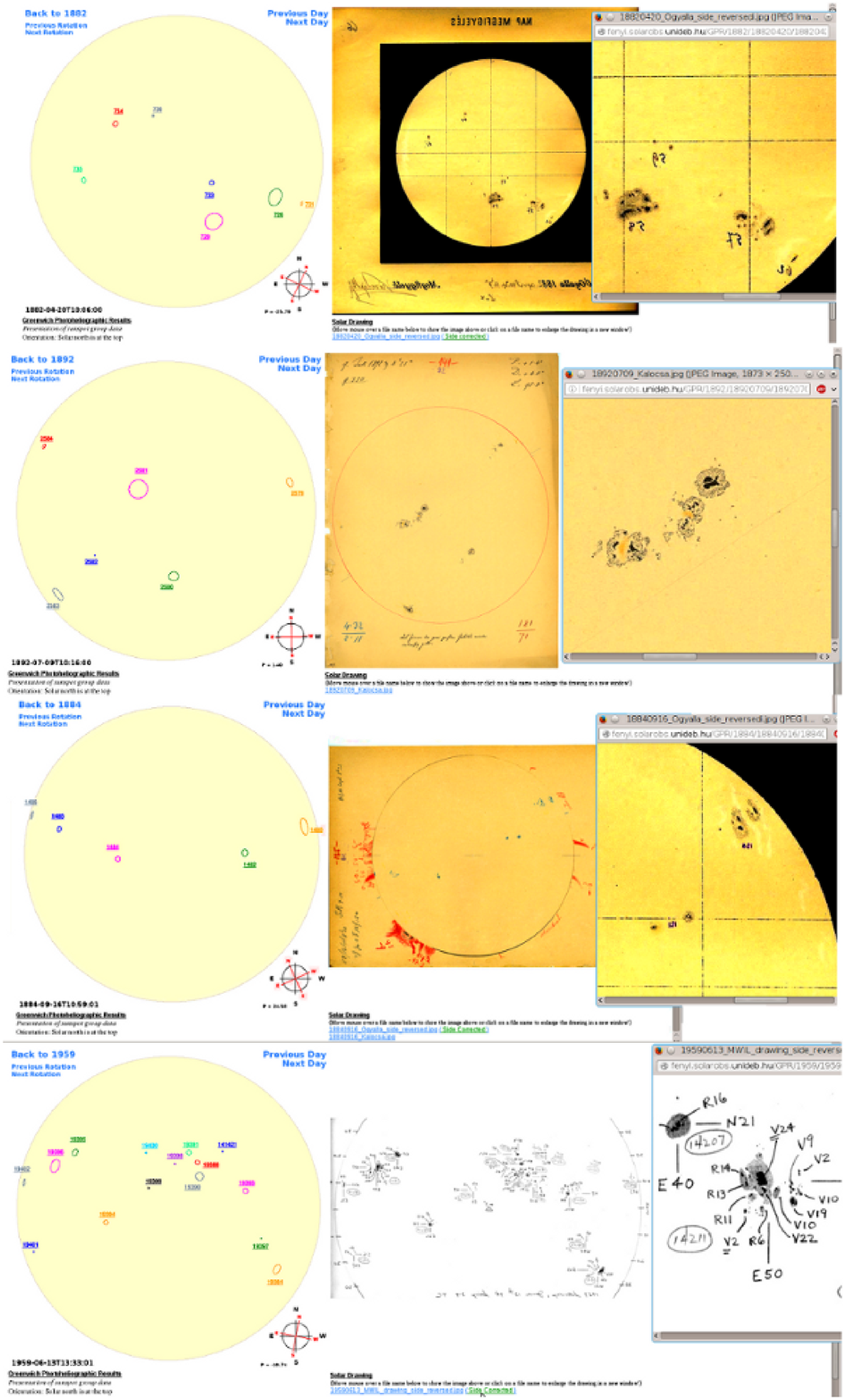}
\caption{Examples for web pages of GPR available at {\sf fenyi.solarobs.unideb.hu/GPR/}. See the detailed explanation in the text.}
\end{center}
\end{figure}

We documented the changes at the page of ``List of data modifications" ({\sf fenyi.solarobs.unideb.hu/GPR/modifications/}) where both the original and the new data can be seen. The improved ``saf" files are also available at the ftp site of DHO ({\sf ftp://fenyi.solarobs.unideb.hu:2121/pub/GPR/saf/}) in the same format as they are created at NGDC. The transitional versions of these files (``FIN.tmp", ``FIN.new") are also published here ({\sf ftp://fenyi.solarobs.unideb.hu\\:2121/pub/GPR/FIN/}), which may help in, {\it e.g.}, handling the temporal data in different formats or the lines of individual spot data in case they are needed. The revised version of ``gpr" files were converted into DPD format. As a result, there are no individual spot data in the rows of sunspot data in the converted files; only the rows of groups are repeated in them to follow the DPD format. These converted files are used to create the web pages of GPR at {\sf fenyi.solarobs.unideb.hu/GPR/}. 

The graphical on-line presentation of the data is completed with the solar drawings of HHSD and the Mount Wilson Observatory. Figure 11 shows four screenshots showing the possible use of the interactive tool of GPR. 

The first snapshot shows a page with a drawing from \'Ogyalla. The  \'Ogyalla drawings are East-West side-reversed as it is earlier mentioned, thus two options are available for displaying them. The original view is suitable for reading the text in the image (date or group number). The side-corrected view is suitable for comparing the GPR data.  If someone moves the mouse over the file names below the image, the versions of the image can be swapped.  If one clicks on a file name, the image opens in a pop-up window in which the image can be enlarged (see the right-hand side of the first snapshot reported in Figure 11).  
If someone wants to compare the observation with the schematic drawing of GPR data, it has to be taken into account that the orientation of the original observation differs from that of the schematic drawing.  In the \'Ogyalla images, the terrestrial North is at the top while in the schematic GPR drawings the solar North is at the top. The small orientation figure  in the middle of the page between the original and  schematic drawings helps in comparing the orientation of the two images showing the position angle $P_{0}$ of the solar North (N in red) measured eastwards from the terrestrial North point (N in black) of the solar disc.

In the second screenshot, a web page can be seen with a drawing from Kalocsa. In this case it is somewhat more difficult to compare the original and schematic drawings because the orientation and its marks in the Kalocsa drawings were not standardized.  The terrestrial EW and/or solar EW are usually indicated with lines crossing the disc and/or marks at the limb (the relative position of the terrestrial and solar coordinate systems is the same as that of the black and red ones in the small orientation drawing but they may be rotated with a quantifiable angle in the observational drawing). If the orientation is indicated ambiguously in a drawing, it can be reconstructed by comparing the various data presented in the related web page.

If there are two original observations for a day, both of them are listed with links to the images in the web page. The third screenshot shows that the user may select two different images to be shown in the main page and in the pop-up window. In this screenshot there is a Kalocsa drawing in the main page and there is an \'Ogyalla drawing in the pop-up window.

The fourth screenshot shows a page with a MW polarity drawing. The MW drawings are also East-West side-reversed observations, thus the correct position of spots and the readable text of magnetic information cannot be seen at the same time. The screenshot shows that the user may select two different views of a drawing to be shown  in the main page and in the pop-up window.

\section{Brief History of Photoheliographic Databases GPR and DPD}

Now that the GPR and DPD data are unified, it is edifying to summarize briefly their common history by using historical resources \citep{Scott85, W87, R93, M75, D87}, the volumes of GPR, a few recent documents (e. g. volumes of IAU Transactions) and personal communications:

1843:
The German amateur astronomer Heinrich Schwabe discovered the 11-year cycles of solar activity.

1845:
The first clear image of the Sun was a daguerreotype taken by A.H.L. Fizeau and J.B.L. Foucault.

1852: 
Edward Sabine announced that the Schwabe's sunspot cycle was correlated very closely with the Earth's 11-year geomagnetic cycle. Astronomers became interested in observing the Sun. 

1854: 
John Herschel argued the importance of obtaining daily photographic pictures of the Sun's disc, the Kew Observatory Committee of the British Association took the matter up, the Royal Astronomical Society decided to support the building a photoheliograph for Kew.

1857: 
Warren De la Rue produced the design for the Kew Photoheliograph, the first telescope specifically built to photograph the Sun.

1858: 
The systematic photographic observations started at Kew Observatory, where Sabine controlled the geomagnetic and meteorological research, and he secured funds for the sunspot record.

1859: 
On 1 September, the magnetometers at Kew recorded a brief but very noticeable jump in the Earth's magnetic field at exactly the same time as two amateur astronomers, R.C. Carrington and R. Hodgson, were the first to observe a flare on the Sun. It was the first observation of a space-weather event.

1860: 
The photoheliograph was briefly removed from Kew to a site in Spain, where De la Rue used it to take the first good pictures of a total solar eclipse. 

1861--72: 
The photoheliograph returned to Kew, where the observers gathered 2778 white-light full-disc photographic observations for a full solar cycle. The observatory gained renown and was well regarded for its three-fold activities (solar physics, geomagnetism, and meteorology). George B. Airy, who was the Astronomer Royal at the Royal Observatory in Greenwich (RGO), regarded Kew as a rival over this decade. Finally, Airy achieved the transfer of Kew's photoheliograph to Greenwich in August 1872.

1873: 
The daily photoheliographic observations and their evaluation started at RGO. The goal was to construct a homogeneous and precise solar dataset and to provide information on solar activity for the Magnetic Observatory of RGO.

1874--1913: 
It was the ``golden era" of GPR based on new instruments, increasing network of contributing observatories, and publications of detailed sunspot and facular data in addition to sunspot group data, and new types of tables ({\it e.g.}for recurrent sunspot groups). The data were published within one to three years after observations. 

1914--36: 
Because of the First World War, the  backlog of publication increased to four to five years. In 1916, the observers decreased the published information: the daily data only contained the mean position and the whole area of sunspot groups but sunspot data were no longer published. This resulted in a quicker publication after 1921; the data were published within one to two years again. 

1937--66: 
The Second World War caused several problems in data production starting with the data for 1937. After the war, the photoheliograph was moved to Herstmonceux from Greenwich to achieve better seeing conditions. These two things together caused a large lack in various resources. Thus, the publication of the GPR suffered from a large delay (9--14 years, average: 12 years). To decrease the time-lag, the observers decreased the information content of GPR again, starting at 1956 the daily detailed facular data were not published any longer and the tables of sunspot groups with projected area data were omitted. After that, the delay of publication was three to nine years (average: 5.9 years).

1967--76: 
In 1967, the Magnetic Observatory became officially separated from RGO. In this way, the direct interest of RGO in solar observations based on research of the relationship between solar activity and geomagnetic storms ceased. However, at this time the publication of GPR was regarded as an international duty of RGO; thus, the GPR project was continued but it had only a very low priority. In 1971, the fundamental reorganization of RGO and the rearrangement of resources to research groups at universities started. Because of these changes, the RGO had to decide to finish a number of its research programs.

1976:
The termination of GPR was announced at the IAU General Assembly (GA) in Grenoble; Commission 10 (Solar activity) accepted this decision. The last volume of GPR for 1972--76 was published in 1980. The DHO of the Hungarian Academy of Sciences  (HAS) had already a long tradition of high-quality photoheliographic observations and precise measurement of sunspot position at this time. Commission 10 encouraged the DHO to undertake this task.

1977--78:
The DHO and the HAS formally assumed responsibility for this program in January 1977 and established collaborative work with Pulkovo, Kislovodsk, Kodaikanal, and RGO to ensure a continuous daily sunspot record.

1979--81:
The Debrecen photoheliograph program was approved at the IAU XVII GA in 1979. The team of Debrecen Photoheliographic Results (DPR) wanted to provide information on sunspots similar to or even exceeding the great content of GPR in the ``golden era". The planned procedure was very time consuming.

1982--1992:
In 1982, the DHO became a department of the Konkoly Observatory (KO), Budapest. The DPR1977 was only published in 1987 \citep{Dea87} and DPR1978 in 1995.

1993--2014:
The solar community encouraged DHO to speed up the process and therefore a separate project was launched to produce the DPD. Its team has left out the most time-consuming components of the DPR-procedure, thus the information content of DPD got closer to that of the early GPR but it was still much more detailed, while the speed of publication of DPD gradually increased. 42 volumes of DPD have been published during 23 years. The revised version of GPR has been converted to DPD format and they were published in a unified form.

2015:
It was announced at the IAU XXIX GA that the authors of the present article, the permanent members of the DPD team, fulfilled the undertaking on sunspot database with more extended content and services than was required.
 
2016:
At the beginning of the year, the director of the Konkoly Observatory decided to close the DHO and its Gyula Observing Station by the end of the year. The future of the DPD catalog is unclear at present but some kind of continuation of the catalog work is planned in the framework of the Konkoly Observatory. (This may mean that the links published in this article will change in the future.)

\section{Examples for the Exploitation of the Databases}

It may be useful to demonstrate the research potential of the databases. Their unique features enabled several recent studies that have been carried out by exploiting these advantages.

Data of both sunspots and sunspot groups in the same tables enable one to track the internal dynamics of the active regions. A detailed study of \citet{Murakozy14} addresses the variation of the distances of leading--following subgroups, the leading--following asymmetry of compactness, as well as of the rates of development. A possible new activity parameter is proposed by \citet{Murakozy16}; this method also needs the magnetic and areal data of each spot within the groups.

The high temporal resolution in the last two decades is a powerful tool to refine the internal processes of active regions, all of the above-mentioned studies exploit it. 
The internal dynamics of sunspot groups leading to flares has been studied by \citet{korsos} by focusing on the mutual displacements of spots of opposite polarities.  By combining various space-borne spot and flare data, a new tool can be developed to study the position of flares within sunspot groups \citep{Gtool}. The position and size of sunspot groups can be useful to reveal the spatial and temporal distribution of small flares before major flares \citep{Gyenge16a}.

The long homogeneous series of sunspot data is indispensable for studies of long-term processes. This has been exploited by \citet{gyenge} in tracking the migration of solar active longitudes. 
After comparing the positions of flares and active longitudes, \citet{Gyenge16b} showed that the most flare-productive active regions tend to be located in or close to the active longitudinal belt. 
The phase lags of solar hemispheric cycles studied by \citet{Murakozy12} also needed a dataset covering more than a century.

The availability of tilt-angle data has allowed the publication of a number of articles recently. The comparison of tilt angles derived from white-light images with those of magnetograms by \citet{Wang} revealed that the latter include the contribution of facular areas, which tend to result in greater axial inclinations than the adjacent sunspots.  The large amount of tilt data allowed the refinement of Joy's law:  \citet{Baranyi15} showed that the refined diagram of Joy's law presents a plateau in the domain around the mean latitude of the active-region belt. The Debrecen data were used by \citet{Pavai} as reference datasets to obtain a statistical mapping of 
sunspot umbral areas derived from historical solar drawings and they contributed to the validation of the historical tilt-angle data.

\citet{kitchatinov} used the DPD for investigating the North--South asymmetry of the solar dynamo 
by estimating  poloidal-field formation from sunspot data.  \citet{Sokoloff, McClintock14}  examined the anti-Hale groups by means of the DPD website. \citet{Sun} compared the spatial distribution of sunspot butterfly diagram constructed from the DPD sunspot catalog, the DPD tilt angles, and the radial magnetic field to study polar magnetic-field reversal and surface flux transport.

By combining the DPD data with other types of data, \citet{Mordvinov} give examples to demonstrate that the regular polar-field build-up is disturbed by surges of leading polarities that resulted from violations of Joy's law at lower latitudes.
\citet{McClintock16} investigated how tilt angle and footpoint separation of bipolar sunspot regions vary during emergence and decay by using HMIDD.

The Debrecen data helped \citet{Moon} in developing a new method of image patch analysis of sunspots and active regions.
In the studies of the Solar Irradiance Climate Data Record by \citet{Coddington}, the DPD will be used as an independent data source that is accessed for quality assurance of the sunspot darkening.

These examples show that there is a wide range of scientific topics that can be studied in an efficient way
by exploiting the data and tools presented in this article.

\section{Summary}

The article presents an  overview about the most complex ensemble of sunspot databases edited by the Heliophysical Observatory, Debrecen, Hungary. The aim of the team was to compile sunspot datasets containing all relevant data for studies of sunspot activity. The developing observational techniques allowed varying data contents in different time intervals, as summarized in Table 1, but the team endeavored to obtain the maximum information from all types of observations. The users of the datasets can easily carry out investigations of sunspots on large statistical samples. While the earlier types of sunspot datasets only enabled to investigate the  sunspot activity globally, the present databases also make possible to study the internal dynamics of sunspot groups.

The presentation of the data is user friendly and provides several tools to make easy the search and selection. The high-level data are compiled in tables but all involved images of observations are appended so the sunspot development can be tracked by stepping through the consecutive observations, while the images have links to the numerical data by clicking on the images of spots. The search is also facilitated by a mySQL query. Thus the database ensemble is not only the most detailed and complete documentation of the sunspot activity but also the most versatile tool to find the necessary data.

\begin{table}
\caption{The contents of sunspot catalogs available at the Debrecen Observatory}
 \label{catalogs}
\begin{tabular}{lccccc} 
\hline                   

 &  GPR & DPD & SDD & HMIDD  & HHSD  \\
 &  revised & &     &        &     \\
\hline                   

interval   & 1872-1976 & 1974--  & 1996--2010 & 2010--  & 1872--1919 \\
group data              & yes         & yes      & yes         & yes      &  no         \\
spot data               & no         & yes      & yes         & yes      &   no        \\ 
cadence         & 1 day     & 1 day  & ${\approx}$1.5 hours & 1 hour & sporadic daily \\
magnetic data           & no         & no      & yes         & yes      &   no        \\
tilt angle          & no         & yes      & yes         & yes      & no          \\
magnetic tilt angle    & no         & no      & yes         & yes      &  no         \\
faculae             & no         & no      & yes         & yes      &  no         \\
white-light images              & HHSD         & yes      & yes         & yes      & yes         \\
magnetic observations              & MW         & yes      & yes         & yes      & no         \\
HTML platform       & yes         & yes      & yes         & yes      & yes       \\
on-line query       & yes         & yes      & yes         & no      & no       \\
  \hline
\end{tabular}
\end{table}

Table 1 summarizes the main features of the mentioned catalogs to overview the information content of the materials of Debrecen accessible at \\{\sf fenyi.solarobs.unideb.hu/deb\_obs\_en.html}.

We cite the resolution of the IAU (IAU Transactions XVIB, 107, 1977) on the responsibilities of DHO concerning the continuation of GPR:\\
-To carry out direct photographic observations at Debrecen\\
-To organize cooperation between other observatories willing to contribute to such a project\\
-With the assistance of the Greenwich Observatory to ensure a homogeneous  continuity of the gathering, reduction and publication of such data\\
-To ensure the archiving of the original photographs and this access to interested scientists from around the world.

The present article shows that the existing materials exceed the above requirements. 
Now the GPR and the DPD constitute a homogeneous dataset that is supported by various the on-line tools. The Debrecen catalogs provide a new window of enhanced resolution to the solar activity.

The DPD team would highly appreciate if the users of the tools and data presented in this article acknowledged its long-term efforts by referring to this article and to the article on the method of evaluation by \citet{Gea16} in their publications.

\begin{acknowledgements}

This work was supported by the European Community's Seventh Framework Program (FP7 SP1-Cooperation) under grant agreement No. 284461 (EHEROES).
The various tasks related to the databases and tools described in this paper were supported during the past 23 years by the following grants:
European Community's Seventh Framework Programme (FP7/2007--2015) under grant agreement No. 284461 (EHEROES, Mar. 2012 -- Feb. 2015) and  No. 218816 (SOTERIA, Nov. 2008 -- Oct. 2011);
ESA PECS contracts No. 98017 (2004--2007) and No. C98081 (2009--2012);
National Development Agency under grant agreement No. BONUS\_HU\_08/2009-003 (2010--11) and T\'AMOP 4.2.2.C-11/1/KONV/2012--0015 (2012--13);
U.S.-Hungarian Joint Fund for Science and Technology under contract No. 95a-524 (1996--1998); 
SCOSTEP supplemental STEP grant (1995); 
Hungarian Ministry of Cultural Heritage under Millenium Program grant agreement No. Sz\"OP422 (1999--2001);
Grants of the Hungarian National Foundation for Scientific Research Nos. OTKA T037725 (2002--2005), T025640 (1998--2000),  T019829 (1996--1999), T014036 (1994--1996), T007422 (1993--1996), F4142 (1992--1995), P31104 (1998), and U21342 (1996).
We thank the referees and grant providers for supporting our proposals.

We express our deepest gratitude to the colleagues at the collaborating observatories for participating in the daily routine observations and putting the necessary material at our disposal. The contributing observatories taking white-light full-disc and/or magnetic observations were: Abastumani Astrophysical Observatory (Georgia), Astronomical Observatory of Ural State University (Russia), INAF-Catania Astrophysical Observatory (Italy), Crimean Astrophysical Observatory (Russia), Ebro Observatory (Spain), Helwan Observatory (Egypt), Huairou Solar Observing Station of National Astronomical Observatories of CAS (China), Institute of Geophysics and Astronomy of Cuba (Cuba), Kanzelh\"ohe Solar Observatory (Austria), Kiev University Observatory (Ukraine), Pulkovo Observatory and its Kislovodsk Observing Station (Russia),  Kodaikanal Observatory (India), Mauna Loa Solar Observatory (USA), Mount Wilson Observatory (USA), San Fernando Observatory (USA), Solar Observatory of National Astronomical Observatory of Japan 
(Japan), Rome Astronomical Observatory (Italy), Royal Observatory of Belgium (USET data/image of Uccle/Brussels, Belgium), Royal Greenwich Observatory (UK), Sayan Observatory of Institute of Solar-Terrestrial Physics of Siberian Department of RAS (Russia), Tashkent Observatory (Uzbekistan), Ussuriysk Astrophysical Observatory of Far-Eastern Branch of the RAS (Russia), Vala\v ssk\'e Mezi\v r\'i\v c\'i Observatory (Czech Republic).

The Mount Wilson white-light full-disc scans are available thanks to the Mt. Wilson Solar Photographic Archive Digitization Project supported by the National Science Foundation (NSF) under Grant No. 0236682. The magnetic database includes data from the synoptic program at the 150-Foot Solar Tower of the Mount Wilson Observatory. The Mt. Wilson 150-Foot Solar Tower is operated by UCLA, with funding from NASA, ONR and NSF, under agreement with the Mt. Wilson Institute.
The observations of Kanzelh\"ohe Solar Observatory are available by courtesy of the Central European Solar ARchives (CESAR).
The {\it Michelson Doppler Imager} (MDI) data are used by courtesy of the SOHO/MDI research group at Stanford University. 
{\it Solar and Heliospheric Observatory} (SOHO) is a mission of international cooperation between ESA and NASA.
The SDO/HMI images are available by courtesy of NASA/SDO and the AIA, EVE, and HMI science teams.
NSO/Kitt Peak magnetic data used here are produced cooperatively by NSF, NASA/GSFC, and NOAA/SEL.
We acknowledge the courtesy of editors of Solnechnie Dannie solar  catalog, who permit the use of magnetic-polarity drawings observed by several contributing observatories.
This work utilizes data obtained by the Global Oscillation Network Group (GONG) Program, managed by the National Solar Observatory, which is operated by AURA, Inc. under a cooperative agreement with the NSF. The data were acquired by instruments operated by the Big Bear Solar Observatory, High Altitude Observatory, Learmonth Solar Observatory, Udaipur Solar Observatory, Instituto de Astrof\'{\i}sica de Canarias, and Cerro Tololo Inter-American Observatory.
Data used here from Mees Solar Observatory, University of Hawaii, are produced with the support of NASA grant NNG06GE13G.
We acknowledge the courtesy of  Yunnan Astronomical Observatory (YNAO) for permitting the use of magnetic-polarity drawings published in Publications of Yunnan Observatory.
The images of Precision Solar Photometric Telescope (PSPT) at Mauna Loa are available by courtesy of the Mauna Loa Solar Observatory, operated by the High Altitude Observatory, as part of the National Center for Atmospheric Research (NCAR). NCAR is supported by the NSF.
We appreciate the long-term work of NOAA/NGDC providing a wide range of scientific products and services for solar physics, and publishing the volumes of Solar-Geophysical Data (SGD).

We thank Norbert Nagy, who was a programmer mathematician at DHO, for playing an important role in development of the on-line tools and data pipeline. We are grateful to our colleagues at DHO and at the collaborating institutes who helped the data evaluation and participated in the observations during the last decades.

\end{acknowledgements}

\section*{Disclosure of Potential Conflicts of Interest}
The authors declare that they have no conflicts of interest.

\bibliographystyle{spr-mp-sola}

\end{article} 

\end{document}